\documentclass[pra,aps,onecolumn,twoside,a4,superscriptaddress,showkeys]{revtex4}

\usepackage{amsfonts}
\usepackage{amsmath}
\usepackage{amssymb}
\usepackage{graphicx}
\usepackage{psfrag}
\usepackage{color}

\definecolor{labelkey}{gray}{.1}
\definecolor{refkey}{gray}{.1}
\definecolor{dgreen}{rgb}{0.000,0.500,0.000}

\begin{document}
\title{\large
\mbox{ Multipartite Entanglement and Hyperdeterminants }}
\author{\large Akimasa Miyake}
\affiliation{ \mbox{Department of Physics, Graduate School of Science, 
University of Tokyo,} \\ 
\mbox{ Hongo 7-3-1, Bunkyo-ku, Tokyo 113-0033, Japan}} 
\affiliation{ \mbox{Quantum Computation and Information Project, 
ERATO, Japan Science and Technology,}  \\
\mbox{ Hongo 5-28-3, Bunkyo-ku, Tokyo 113-0033, Japan}\\
\mbox{\tt Email: miyake@monet.phys.s.u-tokyo.ac.jp, 
            wadati@monet.phys.s.u-tokyo.ac.jp} 
\\ \mbox{} }
\author{\large Miki Wadati}
\affiliation{ \mbox{Department of Physics, Graduate School of Science, 
University of Tokyo,} \\ 
\mbox{ Hongo 7-3-1, Bunkyo-ku, Tokyo 113-0033, Japan}}

%
%
\begin{abstract}
We classify multipartite entanglement in a unified manner, focusing on 
a duality between the set of separable states and that of entangled states.
Hyperdeterminants, derived from the duality, are natural generalizations
of entanglement measures, the concurrence, 3-tangle for 2, 3 qubits 
respectively.
Our approach reveals how inequivalent multipartite entangled classes of pure 
states constitute a partially ordered structure under local actions, 
significantly different from a totally ordered one in the bipartite case.
Moreover, the generic entangled class of the maximal dimension, given by the 
nonzero hyperdeterminant, does not include the maximally entangled states
in Bell's inequalities in general (e.g., in the \(n \!\geq\! 4\) qubits), 
contrary to the widely known bipartite or 3-qubit cases.
It suggests that not only are they never locally interconvertible with
the majority of multipartite entangled states, but they would have no grounds
for the canonical \(n\)-partite entangled states.
Our classification is also useful for that of mixed states.
\end{abstract}
\keywords{multipartite entanglement, hyperdeterminant, duality, 
stochastic LOCC \hfill \\
\mbox{}\\ \mbox{}\\
{\bf
Contribution to the ERATO workshop on Quantum Information Science 2002 
(September 5-8, 2002, Tokyo, Japan), published in Quant. Info. Comp. 
{\bf 2} (Special), 540-555 (2002).} \\
\mbox{}
}

\maketitle

\section{Introduction}               
\label{sec:1}
\noindent
The recent development of quantum information science \cite{review} draws our 
attention to entanglement, the quantum correlation exhibiting nonlocal 
(nonseparable) properties, not only as a useful resource but as a renewed 
fundamental aspect in quantum theory.
Since entanglement is supposed to be never strengthened, on average, by local 
operations and classical communication (LOCC), characterizing it 
under LOCC is one of our basic interests.
Here we classify entanglement of {\it multi}-parties, which is less 
satisfactorily understood than that of {\it two}-parties.

%
%
When we classify the single copy of multipartite pure states on 
the Hilbert space \({\mathcal H}={\mathbb C}^{k_1 +1}\otimes\cdots\otimes
{\mathbb C}^{k_n +1}\) (precisely, rays on its projective space 
\(M={\mathbb C}P^{(k_1 +1)\cdots (k_n +1)-1}\)),
\begin{equation}
\label{eq:multi}
|\Psi\rangle = \sum_{i_1,\ldots,i_n=0}^{k_1,\ldots,k_n} a_{i_1,\ldots,i_n}
|i_1\rangle \otimes \cdots \otimes |i_n\rangle,
\end{equation}
there are many difficulties in applying the techniques, e.g., 
the Schmidt decomposition, utilized in the bipartite case \cite{linden+98}.
Still, we can consider a coarser classification by stochastic LOCC (SLOCC)
\cite{bennett+00,dur+00} than LOCC. 
There we identify two states \(|\Psi\rangle\) and \(|\Phi\rangle\) that 
convert to each other back and forth with (maybe different) nonvanishing 
probabilities, in contrast with LOCC where we identify the states 
interconvertible deterministically.
These states \(|\Psi\rangle\) and \(|\Phi\rangle\) are supposed to perform 
the same tasks in quantum information processing, although their probabilities 
differ.
Later, we find that this SLOCC classification is still fine grained to classify
the multipartite entanglement.
Mathematically, two states belong to the same class under SLOCC if and only if
they are converted by an {\it invertible} local operation \(G\) having a 
nonzero determinant \cite{dur+00}. 
Thus the SLOCC classification is equivalent to the classification of orbits of 
the natural action: direct product of general linear groups 
\(GL_{k_{1}+1}({\mathbb C})\!\times\!\cdots\!\times\! 
GL_{k_{n}+1}({\mathbb C})\) \cite{note1}.

In the bipartite (for simplicity, \(k_1 = k_2 =k\)) case, the SLOCC 
classification means just classifying the whole states \(M\) by the rank of 
the coefficient matrix \(A \!=\! (a_{i_1,i_2})\), also known as the (Schmidt) 
local rank \cite{note2}, because \(A\) is transformed as
\begin{equation}
A \;\;\stackrel{G}{\longrightarrow}\;\; G^{(1)} \; A \; G^{(2)T},
\end{equation}
under an invertible local operation \(G=G^{(1)} \otimes G^{(2)} \in 
GL_{k+1}\times GL_{k+1}\) (the superscript \(T\) stands for the transposition)
so that its rank is the SLOCC invariant.
A set \(S_j\) of states of the local rank \(\leq j\) is a {\it closed} 
subvariety under SLOCC and \(S_{j-1}\) is the singular locus of \(S_j\). 
This is how the local rank leads to an "onion" structure \cite{acin+01} 
(mathematically the stratification):
\begin{equation}
\label{eq:onion}
M\!=\!S_{k+1} \supset S_{k}\supset\cdots\supset S_1 \supset 
S_0 \!=\! \emptyset, 
\end{equation}
and \(S_{j}\!-\! S_{j-1}\;(j=1,\ldots,k\!+\!1)\) give \(k\!+\!1\) classes of 
entangled states.
Now we discuss the relationship between these classes under {\it noninvertible}
local operations.
Since the local rank can decrease by noninvertible local operations, 
i.e., general LOCC \cite{lo+01},
we find that \(k\!+\!1\) classes are totally ordered.
In particular, 
the outermost generic set \(S_{k+1}\!-\!S_{k}\) is the class of maximally 
entangled states, for this class can convert to all classes by LOCC but
other classes never convert to it.
The innermost set \(S_{1}(\!=\! S_{1}\!-\!S_{0})\) is that of separable 
(no-entangled) states. Indeed this class never convert to other classes 
by LOCC but any other classes can convert to it.

In the 3-qubit case, D\"{u}r {\it et al.} showed that SLOCC classifies 
\(M\) into {\it finite} classes and in particular there 
exist two inequivalent, Greenberger-Horne-Zeilinger (GHZ) and W, classes of 
the genuine tripartite entanglement \cite{dur+00}. 
They also pointed out that the SLOCC classification has {\it infinitely} many 
orbits in general (e.g., for \(n\geq 4\)).
In this paper, we classify multipartite entangled states in a unified manner 
based on hyperdeterminants, and clarify how they are partially ordered.
The advantages are three-fold. 
\begin{enumerate}
\item This classification is equivalent to the SLOCC classification when 
SLOCC has finitely many orbits. So it naturally includes the widely known 
bipartite and \(3\)-qubit cases.

\item In the multipartite case, we need further SLOCC invariants in addition
to the local ranks. For example, in the \(3\)-qubit case \cite{dur+00}, 
the \(3\)-tangle \(\tau\), just the absolute value of the hyperdeterminant 
(see Eq.~(\ref{eq:tau})), is utilized to distinguish GHZ and W classes.
This work clarifies why the \(3\)-tangle \(\tau\) appears and how these SLOCC
invariants are related to the hyperdeterminant in general.

\item Our classification is also useful to multipartite mixed states.
A mixed state \(\rho\) can be decomposed as a convex combination of 
projectors onto pure states.
Considering how \(\rho\) needs at least the outer class in the onion 
structure of pure states, we can also classify multipartite mixed states
into the totally ordered classes (for details, see the 
appendix of \cite{miyake02}). We concentrate the pure states here.
\end{enumerate}

%
%
The sketch of our idea is as follows.
We focus on a duality between the set of separable states and
the set of entangled states.
The set of completely separable states is the smallest closed 
subvariety, called Segre variety, \(X\), while its dual variety \(X^{\vee}\) 
is the largest closed subvariety which consists of {\it degenerate} 
entangled states (precisely, if \(X^{\vee}\) is 1-codimensional).
Indeed, in the bipartite (\(k_1 = k_2 =k\)) case, it means that \(X\) is 
the set of states of the local rank \(1\), i.e., \(X=S_1\).
On the other hand, \(X^{\vee}\) is the set of states 
where the local rank is not full (\(\det A \!=\!0\)), i.e., \(X^{\vee}=S_k\). 
The duality between the smallest subvariety \(X\) and the largest subvariety
\(X^{\vee}\) holds also for the multipartite case (e.g., see 
Fig.~\ref{fig:onion_3bit}), and 
the dual variety \(X^{\vee}\) is given, in analogy, by the zero 
hyperdeterminant: \({\rm Det}A \!=\!0\). 
The outside of \(X^{\vee}\), i.e., \({\rm Det}A \!\ne\! 0\), is the 
generic (non degenerate) entangled class, and \(|{\rm Det}A|\) is 
the entanglement measure which represents the amount of generic entanglement.
It is also known as the concurrence \(C\) \cite{hill+97}, 
\(3\)-tangle \(\tau\) \cite{coffman+00} for the \(2,3\)-qubit pure case, 
respectively (see Sec.3).
It is significant that \({\rm Det}A\) is relatively invariant under SLOCC.
In order to obtain other (degenerate) entangled classes, we need to decide the
singularities of \(X^{\vee}\) as we did in the bipartite case.
After this onion-like classification of entangled classes (SLOCC orbits),
we characterize the relationship between them under noninvertible local 
operations.
This reveals how multipartite entangled classes are partially ordered,
contrary to the bipartite case.
We clarify what this structure looks like, as the dimensions 
\(k_j \!+\! 1\) of subsystems become larger, or as the number \(n\) of 
the parties increases.

%
%
Accordingly, the rest of the paper is organized as follows.
In Sec.2, the duality between separable states and entangled states is 
introduced. 
The hyperdeterminant, associated to this duality, and its singularities lead 
to the SLOCC-invariant onion-like structure of multipartite entanglement.
The characteristics of the hyperdeterminant and its singularities are 
explained in Sec.3.
Classifications of multipartite entangled states are exemplified in 
Sec.4 so as to reveal how they are ordered under SLOCC. 
Finally, the conclusion is given in Sec.5.

\section{Duality between separable states and entangled states}
\label{sec:2}
\noindent
In this section, we find that there is a duality between the set of 
separable states and that of entangled states.
This duality derives the hyperdeterminant our classification is 
based on.

\subsection{Preliminary: Segre variety}
\label{sec:segre}
\noindent
%
%
%
To introduce our idea, we first recall the geometry of pure states.
In a complex (finite) \(k\!+\!1\)-dimensional Hilbert space 
\({\mathcal H}({\mathbb C}^{k+1})\), let \(|\Psi\rangle\) be a (not 
necessarily normalized) vector given by \(k\!+\!1\)-tuple of complex 
amplitudes \(x_{j}(j=0,\ldots,k)\in{\mathbb C}^{k+1}\!-\!\{0\}\) in 
a computational basis (i.e., \(x_j\) are the coefficients in 
Eq.~(\ref{eq:multi}) for \(n=1, k_1=k\)).
The physical state in \({\mathcal H}({\mathbb C}^{k+1})\) is a ray, 
an equivalence class of vectors up to an overall nonzero complex number.
Then the set of rays constitutes the complex projective space 
\({\mathbb C}P^{k}\) and \(x := (x_{0}:\ldots : x_{k})\), considered up to 
a complex scalar multiple, gives homogeneous coordinates in 
\({\mathbb C}P^{k}\) \cite{harris}.

%
%
For a composite system which consists of 
\({\mathcal H}({\mathbb C}^{k_{1}+1})\) and 
\({\mathcal H}({\mathbb C}^{k_{2}+1})\),
the whole Hilbert space is the tensor product 
\({\mathcal H}({\mathbb C}^{k_{1}+1})\otimes
{\mathcal H}({\mathbb C}^{k_{2}+1})\)
and the associated projective space is 
\(M={\mathbb C}P^{(k_{1}+1)(k_{2}+1)-1}\).   
A set \(X\) of the separable states is the mere Cartesian product
\({\mathbb C}P^{k_{1}}\times{\mathbb C}P^{k_{2}}\), whose dimension 
\(k_{1}\!+\!k_{2}\) is much smaller than that of the whole space \(M\),
\((k_{1}\!+\!1)(k_{2}\!+\!1)\!-\!1\).
This \(X\) is a closed, smooth algebraic subvariety (Segre variety) defined by 
the Segre embedding into \({\mathbb C}P^{(k_{1}+1)(k_{2}+1)-1}\) 
\cite{harris,miyake+01},
\begin{gather}
{\mathbb C}P^{k_{1}}\times {\mathbb C}P^{k_{2}} \hookrightarrow
{\mathbb C}P^{(k_{1}+1)(k_{2}+1)-1} \nonumber \\
\begin{align}
&\left((x^{(1)}_0:\ldots :x^{(1)}_{k_1}),
(x^{(2)}_0:\ldots :x^{(2)}_{k_2})\right)  \\
&\mapsto\; (x^{(1)}_0 x^{(2)}_0 : \ldots : x^{(1)}_0 x^{(2)}_{k_2} : 
x^{(1)}_1 x^{(2)}_0 : \ldots\ldots : x^{(1)}_{k_1} x^{(2)}_{k_2}). \nonumber
\end{align}
\end{gather}
Denoting homogeneous coordinates in \({\mathbb C}P^{(k_{1}+1)(k_{2}+1)-1}\)
by \(b_{i_1,i_2}\!=\! x^{(1)}_{i_1}x^{(2)}_{i_2} \; (0\leq i_j \leq k_j)\), 
we find that the Segre variety \(X\) is given by the common zero locus of 
\(k_{1}(k_{1}\!+\!1)k_{2}(k_{2}\!+\!1)/4\) 
homogeneous polynomials of degree \(2\):
\begin{equation}
b_{i_{1},i_{2}} b_{i'_{1}, i'_{2}} - b_{i_{1},i'_{2}} b_{i'_{1}, i_{2}},
\end{equation}
where \(0 \leq i_{1} < i'_{1} \leq k_1, \; 0 \leq i_{2} < i'_{2} \leq k_2\).
Note that this condition implies that all \(2\times 2\) minors of the "matrix" 
\(B \!=\! (b_{i_1,i_2})\) equal 0; i.e., the rank of \(B\) is 1.
Thus we have \(X=S_1\), which agrees with the SLOCC classification by 
the local rank in the bipartite case.

%
%
Now consider the multipartite Cartesian product 
\(X\!=\!{\mathbb C}P^{k_{1}}\!\times\!\cdots\!\times\!{\mathbb C}P^{k_{n}}\) 
in the Segre embedding into 
\(M\!=\!{\mathbb C}P^{(k_{1}+1)\cdots(k_{n}+1)-1}\).
Because this Segre variety \(X\) is (the projectivization of) the variety 
composed of the matrices \(B \!=\! (b_{i_1,\ldots, i_n}) \!=\! 
(x^{(1)}_{i_1}\cdots x^{(n)}_{i_n})\), 
it gives a set of the completely separable states
in \({\mathcal H}({\mathbb C}^{k_{1}+1})\!\otimes\cdots\otimes\! 
{\mathcal H}({\mathbb C}^{k_{n}+1})\). 
By another Segre embedding, say \(X'\!=\!{\mathbb C}P^{(k_{1}+1)(k_{2}+1)-1}
\!\times\!{\mathbb C}P^{k_3}\!\times\cdots\times\!{\mathbb C}P^{k_n}\), 
we also distinguish a set of separable states where only 1st and 2nd 
parties can be entangled, i.e., when we regard 1st and 2nd parties as 
one party, an element of this set is completely separable for 
"\(n\!-\!1\)" parties.
This is how, also in the multipartite case, we can classify all kinds of 
{\it separable} states, typically lower dimensional sets.
Note that, in the multipartite case, this check for the separability is 
stricter than the check by local ranks \cite{note4}.

\subsection{Main idea: duality}
\label{sec:dual}
\noindent
We rather want to classify {\it entangled} states, typically higher 
dimensional complementary sets of separable states.
Our strategy is based on the duality in algebraic geometry \cite{harris}; 
a hyperplane in \({\mathbb C}P\) forms the point of a dual projective space 
\({\mathbb C}P^{\ast}\), and conversely every point \(p\) of \({\mathbb C}P\) 
is tied to a hyperplane \(p^{\vee}\) in \({\mathbb C}P^{\ast}\) as 
the set of all hyperplanes in \({\mathbb C}P\) passing through \(p\).
Let us identify the space of \(n\)-dimensional matrices with its dual by 
means of the pairing,
\begin{equation}
\label{eq:F}
F(A,B) = \sum_{i_1,\ldots, i_n =0}^{k_1,\ldots,k_n} 
a_{i_1,\ldots, i_n} b_{i_1,\ldots, i_n}
\end{equation}
(Although in quantum mechanics we take the complex conjugate 
\(a^{\ast}_{i_1,\ldots,i_n}\), {\it compared with \(b_{i_1,\ldots,i_n}\)}, 
for the product, it does not matter here and we avoid writing the unnecessary 
superscript.).
For a state whose homogeneous coordinates are given by \(A\) in 
\({\mathbb C}P^{\ast}_{A}\),
\(F(A,B)=0\) uniquely determines the hyperplane in \({\mathbb C}P_{B}\),
which consists of its orthogonal states.
Conversely, any hyperplane \(F(A,B)=0\) in \({\mathbb C}P_{B}\) gives 
one-to-one correspondence to the point in \({\mathbb C}P^{\ast}_{A}\) by
its coefficients \(A\).
This is the duality between points and hyperplanes.

\begin{figure}[tb]
\begin{center}
\begin{psfrags}
\psfrag{X}{\color{dgreen} {\large $X$}}
\psfrag{x}{\normalsize $x$}
\psfrag{F}{\color{blue}\normalsize $F(A,B)=0$}
\psfrag{V}{\color{red} {\large $X^{\vee}$}}
\psfrag{D}{\color{red}\normalsize ${\rm Det}A =0$ }
\psfrag{B}{\normalsize $B$}
\psfrag{A}{\normalsize $A$}
\includegraphics[width=10cm,clip]{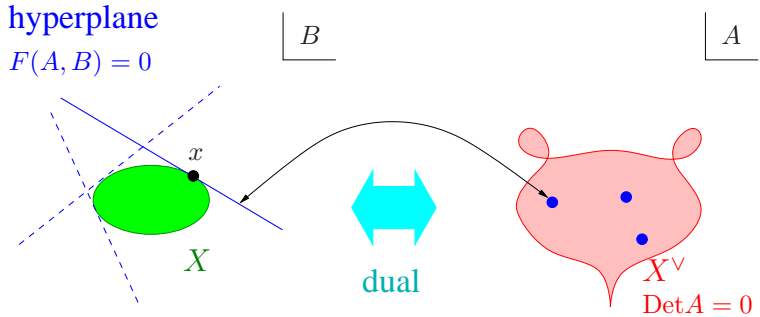}
\caption{The duality between the Segre variety \(X\) and its dual variety
\(X^{\vee}\). The set of all hyperplane tangent to \(X\) in 
\({\mathbb C}P_{B}\) constitute \(X^{\vee}\) in \({\mathbb C}P^{\ast}_{A}\).} 
\label{fig:dual}
\end{psfrags}
\end{center}
\end{figure}

Remarkably, the projective duality between projective subspaces, like the 
above example, can be extended to an involutive correspondence between 
irreducible algebraic subvarieties in \({\mathbb C}P\) and 
\({\mathbb C}P^{\ast}\).
We define a projectively dual (irreducible) variety \(X^{\vee} \subset 
{\mathbb C}P^{\ast}\) as the closure of the set of all hyperplanes tangent 
to the Segre variety \(X\) (see Fig.~\ref{fig:dual}).
As sketched in Sec.1, let us observe (and see the reason later) that, 
for the bipartite case, the variety \(S_{k}\) of the {\it degenerate} 
\((k\!+\!1) \times (k\!+\!1)\) matrices \(A \!=\! (a_{i_1,i_2})\) is 
projectively dual to the variety \(S_1 \!=\! X\) of the matrices 
\(B \!=\! (b_{i_1,i_2}) \!=\! (x^{(1)}_{i_1} x^{(2)}_{i_2})\).
That is, \(S_{k}\) is the dual variety \(X^{\vee}\). 
Following an analogy with a 2-dimensional (bipartite) case, 
an \(n\)-dimensional matrix \(A \!=\! (a_{i_1,\ldots,i_n})\) is called 
{\it degenerate} if and only if it (precisely, its projectivization) lies 
in the projectively dual variety \(X^{\vee}\) of the Segre variety \(X\).
In other words, \(A\) is degenerate if and only if  
its orthogonal hyperplane \(F(A,B) \!=\! 0\) is tangent to \(X\) at some 
nonzero point \(x=(x^{(1)},\ldots, x^{(n)})\) (cf. Fig.~\ref{fig:dual}). 
Analytically, a set of equations, 
\begin{equation}
\left\{\mbox{  }
\begin{split}
\label{eq:critical}
& F(A,x) = \sum_{i_1,\ldots, i_n =0}^{k_1,\ldots,k_n} 
a_{i_1,\ldots, i_n} x^{(1)}_{i_1}\cdots x^{(n)}_{i_n} =  0 \\
& \frac{\partial}{\partial x^{(j)}_{i_j}} F(A,x) = 0 
\quad \mbox{for all} \; j,i_j  
\end{split}
\right. 
\end{equation}
(\(j=1,\ldots,n\) and \(0 \leq i_j \leq k_j\)),
has at least a nontrivial solution \(x \!=\! (x^{(1)},\ldots,x^{(n)})\) of 
every \(x^{(j)} \!\ne\! 0\), and then \(x\) is called a critical point.
The above condition is also equivalent to saying that 
the kernel \({\rm ker} F\) of \(F(A,x)\) is not empty,
where \({\rm ker} F\) is the set of points 
\(x=(x^{(1)},\ldots,x^{(n)}) \in X\) such that, in every \(j_0 = 1,\ldots, n\),
\begin{equation}
F \mbox{\mathversion{bold}$($}
A,(x^{(1)},\ldots,x^{(j_0 -1)},z^{(j_0)},x^{(j_0 +1)},\ldots,x^{(n)})
\mbox{\mathversion{bold}$)$}=0,
\end{equation} 
for the {\it arbitrary} \(z^{(j_0)}\).

In the case of \(n\!=\!2\), the condition for Eqs.(\ref{eq:critical}) 
coincides with the usual notion of degeneracy and 
means that \(A \!=\! (a_{i_1,i_2})\) does not have the full rank.
It shows that \(X^{\vee}\) is nothing but \(S_{k}\).  
In particular, \(X^{\vee}\), defined by this condition, is of codimension 1 
and is given by the ordinary determinant \(\det A \!=\! 0\), if and only if 
\(A\) is a square \((k_{1} \!=\! k_{2} \!=\! k)\) matrix.
In the \(n\)-dimensional case, if \(X^{\vee}\) is a hypersurface 
(of codimension 1), it is given by the zero locus of a unique (up to sign) 
irreducible homogeneous polynomial over \({\mathbb Z}\) of 
\(a_{i_{1},\ldots,i_{n}}\).
This polynomial is the hyperdeterminant introduced by Cayley \cite{cayley} 
and is denoted by \({\rm Det} A\).
As usual, if \(X^{\vee}\) is not a hypersurface, we set \({\rm Det} A\)
to be 1. 

Remember that, in the {\it bipartite} case, 
we classify the states \(\in S_{k+1}\!-\!S_{k} \!=\!  M\!-\!X^{\vee}\) as 
the generic entangled states, the states 
\(\in S_{k}\!-\!S_{k-1} \!=\!  X^{\vee}\!-\!X^{\vee}_{\rm sing}\) as the 
next generic entangled states, and so on.
Likewise, we aim to classify the {\it multipartite} entangled states into the 
onion structure by the dual variety \(X^{\vee}\) 
(\({\rm Det}A\!=\!0\)), its singular locus \(X^{\vee}_{\rm sing}\) and so on, 
i.e., by every closed subvariety.

\section{Hyperdeterminant and its singularities }
\label{sec:3}
\noindent
In order to classify multipartite entanglement into the SLOCC-invariant onion 
structure, we explore the dual variety \(X^{\vee}\) (zero hyperdeterminant) 
and its singular locus in this section.

\subsection{Hyperdeterminant} 
\label{sec:hdet}
\noindent
We utilize the hyperdeterminant, the generalized determinant for 
higher dimensional matrices by Gelfand {\it et al.}
\cite{gelfand+92,gelfand+94}.
Its absolute value is also known as an entanglement measure,
the concurrence \(C\) \cite{hill+97}, 3-tangle \(\tau\) \cite{coffman+00} 
respectively, for the 2,3-qubit pure case.
\begin{align}
\label{eq:C}
C &= 2|{\rm Det} A_2| = 2|\det A|= 2|a_{00}a_{11}-a_{01}a_{10}|, \\
\label{eq:tau}
\tau &= 4|{\rm Det}A_3| \nonumber\\
     &= 4| a_{000}^2 a_{111}^2 + a_{001}^2 a_{110}^2 +
        a_{010}^2 a_{101}^2 + a_{100}^2 a_{011}^2  
        -2(a_{000}a_{001}a_{110}a_{111}+a_{000}a_{010}a_{101}a_{111}\nonumber\\
&\quad\mbox{} 
        + a_{000}a_{100}a_{011}a_{111}+a_{001}a_{010}a_{101}a_{110} 
        + a_{001}a_{100}a_{011}a_{110}+a_{010}a_{100}a_{011}a_{101})\nonumber\\
&\quad\mbox{} 
        + 4 (a_{000}a_{011}a_{101}a_{110} + a_{001}a_{010}a_{100}a_{111})|.
\end{align}

%
%
The following useful facts are found in \cite{gelfand+94}.
Without loss of generality, we assume that \(k_{1}\!\geq\! k_{2} \!\geq\!
\cdots\!\geq\! k_{n} \!\geq\! 1\).
The \(n\)-dimensional hyperdeterminant \({\rm Det} A\) of format 
\((k_{1}\!+\!1)\!\times\!\cdots\!\times\!(k_{n}\!+\!1)\) exists, 
i.e., \(X^{\vee}\) is a hypersurface, if and only if a "polygon inequality"
\(k_1 \!\leq\! k_2 + \cdots\!+\! k_n\) is satisfied. 
For \(n\!=\!2\), this condition is reduced to \(k_{1} \!=\! k_{2}\) 
as desired, and \({\rm Det} A\) coincides with \(\det A\). 
The matrix format is called boundary if \(k_{1} \!=\! k_{2} \!+ \cdots 
+\! k_{n}\) and interior if \(k_{1} \!<\! k_{2} \!+\cdots+\! k_{n}\).
Note that (i) The boundary format includes the "bipartite cut" between 
1st parties and the others so that it is mathematically tractable. 
(ii) The interior format includes the \(n \!\geq\! 3\)-qubit case.
We treat hereafter the format where the polygon inequality holds and  
\(X^{\vee}\) is the largest closed subvariety, defined by the hypersurface 
\({\rm Det}A =0\).

\({\rm Det} A\) is relatively invariant (invariant up to constant) under the 
action of 
\(GL_{k_1 + 1}({\mathbb C})\times\cdots\times GL_{k_n + 1}({\mathbb C})\).
In particular, interchanging two parallel slices (submatrices with some fixed 
directions) leaves \({\rm Det} A\) invariant up to sign, and \({\rm Det} A\) 
is a homogeneous polynomial in the entries of each slice. 
Since it is ensured that \(X^{\vee}\), \(X^{\vee}_{\rm sing}\) and further
singularities are invariant under SLOCC, our classification is equivalent to 
or coarser than the SLOCC classification. Later, we see that the former and 
the latter correspond to the case where SLOCC gives finitely and infinitely 
many classes, respectively.

\subsection{Schl\"{a}fli's construction}
\label{sec:schlafli}
\noindent
It would be not easy to calculate \({\rm Det}A\) directly by its definition 
that Eqs.(\ref{eq:critical}) have at least one solution.
Still, the Schl\"{a}fli's method enables us to construct \({\rm Det}A_n\) 
of format \(2^n\) (\(n\) qubits) by induction on \(n\) 
\cite{gelfand+92,gelfand+94,schlafli}.

For \(n\!=\!2\), by definition \({\rm Det}A_{2} \!=\!\det A \!=\!
a_{00}a_{11}\!-\!a_{01}a_{10}\).
Suppose \({\rm Det} A_{n}\), whose degree of homogeneity is \(l\), is given.
Associating an \(n\!+\!1\)-dimensional matrix 
\(A \!=\! (a_{i_0,i_1,\ldots,i_n})\)  (\(i_{j}\!=\!0,1\)) to a family 
of \(n\)-dimensional matrices \(\widetilde{A}(x) \!=\! 
\left(\sum_{i_0}a_{i_{0},i_{1},\cdots ,i_{n}}x_{i_0}\right) \) 
which linearly depend on the auxiliary variable \(x_{i_0}\),
we have \({\rm Det} \widetilde{A}(x)_{n}\).
Due to Theorem 4.1, 4.2 of \cite{gelfand+94}, the discriminant \(\Delta\) of 
\({\rm Det} \widetilde{A}(x)_{n}\) gives \({\rm Det} A_{n+1}\) with an extra
factor \(R_n\).
The Sylvester formula of the discriminant \(\Delta\) for binary forms enables 
us to write \({\rm Det}A_{n+1}\) in terms of the determinant of order 
\(2l\!-\!1\);
\begin{equation}
\label{eq:schlafli}
{\rm Det}A_{n+1}
=\frac{\Delta ({\rm Det} \widetilde{A}(x)_{n})}{R_{n}} 
=\frac{1}{R_n c_l} \left| \begin{array}{cccccccc}
c_0    & c_1   &\cdots& c_{l-2}  & c_{l-1} & c_l    &\cdots& 0\\
0      & c_0   &\cdots& \cdots   & c_{l-2} & c_{l-1}&\cdots& 0\\
\vdots &       &\ddots&          & \vdots  &        &\ddots& \vdots\\
0      & 0     &\cdots&  c_0     & c_1     & \cdots &\cdots& c_l \\
1 c_1  & 2 c_2 &\cdots& \cdots   & l c_l   & 0      &\cdots& 0\\
\vdots &       &\ddots&          & \vdots  &        &\ddots& \vdots\\
0      & 0     &\cdots& 0        & 1 c_1   & 2 c_2  &\cdots& l c_l\\ 
\end{array}\right|,
\end{equation}
where each \(c_j\) is the coefficient of \(x_{0}^{l-j} x_{1}^j\) in 
\({\rm Det} \widetilde{A}(x)_{n}\), i.e., 
\(c_j = \frac{1}{(l-j)! \; j! }
\frac{\partial^{l}}{\partial x_{0}^{l-j} \partial x_{1}^{j}} 
 {\rm Det} \widetilde{A}(x)_{n}\). 

Note that because for \(n\!=\! 2, 3\), the extra factor \(R_n\) is just 
a nonzero constant, \({\rm Det}A_{3,4}\) for the \(3, 4\) qubits is readily 
calculated respectively.
It would be instructive to check that \({\rm Det}A_3\) in Eq.(\ref{eq:tau}) 
is obtained in this way.
On the other hand, for \(n\!\geq\! 4\), \(R_n\) is the Chow form (related 
resultant) of irreducible components of the singular locus 
\(X^{\vee}_{\rm sing}\).
These are due to the fact that \(X^{\vee}_{\rm sing}\) has codimension \(2\) 
in \(M\) for any formats of the dimension \(n \!\geq\! 3\) except for 
the format \(2^3\) (\(3\)-qubit case), which was conjectured in 
\cite{gelfand+92} and was proved in \cite{weyman+96}.
So we have to explore \(X^{\vee}_{\rm sing}\) not only to classify entangled 
states in the \(n\) qubits, but to calculate \({\rm Det}A_{n+1}\) inductively.
Although \({\rm Det}A_{n\geq 5}\) has yet to be written explicitly, 
only its degree \(l\) of homogeneity is known 
(in Corollary 2.10 of \cite{gelfand+94}) to grow very fast as 
\(2,4,24,128,880,6816,60032,589312,6384384\) for \(n=2,3,\ldots,10\).
It can be said that this monstrous degree reflects the richness of 
multipartite entanglement, compared with the {\it linear} scaling 
(\(\propto k\)) of the degree along the dimensional direction for the 
bipartite \((k\!+\!1) \times (k\!+\!1)\) case.

\subsection{Singularities of the hyperdeterminant }
\label{sec:sing}
\noindent
We describe the singular locus of the dual variety \(X^{\vee}\).
The technical details are given in \cite{weyman+96}.
It is known that, for the boundary format, the next largest closed subvariety 
\(X^{\vee}_{\rm sing}\) is always an irreducible hypersurface in \(X^{\vee}\);
in contrast, for the interior one, \(X^{\vee}_{\rm sing}\) has generally 
two closed irreducible components of codimension 1 in \(X^{\vee}\), 
{\it node} \(X^{\vee}_{\rm node}\) and {\it cusp} \(X^{\vee}_{\rm cusp}\) type 
singularities.
The rest of this subsection can be skipped for the first reading.
It is also illustrated for the 3-qubit case in the appendix of \cite{miyake02}.

\begin{figure}[tb]
\begin{center}
\begin{psfrags}
\psfrag{X}{\large{$X$}}
\psfrag{V}{\large{$X^{\vee}$}}
\includegraphics[clip]{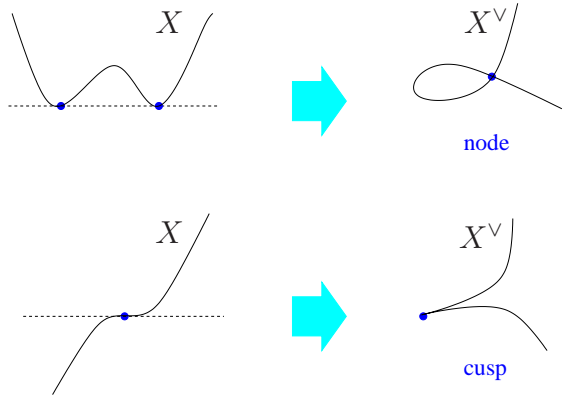}
\caption{
Two types of singularities of \(X^{\vee}\). 
\(X^{\vee}_{\rm node}\) corresponds to the bitangent of \(X\), where
both tangencies are of the first order.
\(X^{\vee}_{\rm cusp}\) corresponds to the tangent at an inflection point 
of \(X\), where its tangency is of the second order. } 
\label{fig:sing_xv}
\end{psfrags}
\end{center}
\end{figure}

First, \(X^{\vee}_{\rm node}\) is the closure of the set of 
hyperplanes tangent to the Segre variety \(X\) at more than one points 
(cf. Fig.~\ref{fig:sing_xv}).
\(X^{\vee}_{\rm node}\) can be composed of closed irreducible subvarieties 
\(X^{\vee}_{\rm node}(J)\) labeled by the subset 
\(J \!\subset\! \{1,\ldots,n\},\) including \(\emptyset\). 
Indicating that two solutions \(x\!=\!(x^{(1)},\ldots,x^{(j)},\ldots,
x^{(n)})\) of Eq.(\ref{eq:critical}) coincide for \(j\!\in\! J\), 
the label \(J\) distinguishes the pattern in these solutions. 
In order to rewrite \(X^{\vee}_{\rm node}(J)\), let us pick up a point 
\(x^o(J)\) such that its homogeneous coordinates 
\(x^{(j)}_{i_j} = \delta_{i_j,0}\) for \(j\!\in\! J\) and 
\(\delta_{i_j,k_j}\) for \(j \!\notin\! J\). 
It is convenient to label the positions of \(1\) in each \(x^{(j)}\) 
by a multi-index \([i_1,\ldots,i_n]\).
For example, \(x^o(1)\) is labeled by \([0,k_2,\ldots,k_n]\) and 
\(x^o(1,\ldots,n)\) is just written by \(x^o\).
When \(X^{\vee}\) is the hyperplane tangent to \(X\) at \(x^o(J)\),
its "\(x^o(J)\)-section" \(X^{\vee}|_{x^o(J)}\) is given as
\begin{equation}
\label{eq:x^o}
X^{\vee}|_{x^o(J)} = 
\left\{ A \; \left| \; \mbox{all \(a_{i'_1,\ldots,i'_n} \!=\! 0\)  s.t. }
\begin{array}{l}
\mbox{\([i'_1,\ldots,i'_n]\) differs from \([i_1,\ldots,i_n]\)} \\ 
\mbox{of \(x^o(J)\) in at most one index} 
\end{array} \right. \right\},
\end{equation}
in order that Eqs.(\ref{eq:critical}) have the nontrivial solution \(x^o(J)\).
Then in terms of the hyperplane bitangent to \(X\) at \(x^o\) and \(x^{o}(J)\),
we can define \(X^{\vee}_{\rm node}(J)\) as  
\begin{equation}
\label{eq:node}
X^{\vee}_{\rm node}(J) =
\overline{(X^{\vee}|_{x^o} \cap X^{\vee}|_{x^o(J)})\cdot G},
\end{equation} 
where \(G\!=\!GL_{k_1+1}\!\times\!\cdots\!\times\! GL_{k_n+1}\) acts \(M\) on 
from the right and the bar stands for the closure.

Second, \(X^{\vee}_{\rm cusp}\) is the set of hyperplanes having a critical 
point which is not a simple quadratic singularity (cf. Fig.~\ref{fig:sing_xv}).
Precisely, the quadric part of \(F(A,x)\) at \(x^o\) is a matrix
\(y_{(j,i_j),(j',i_{j'})}\!=\!
(\partial^2 /\partial x_{i_{j}}^{(j)}\partial x_{i_{j'}}^{(j')})F(A,x^o)\), 
where the pairs \((j,i_j),\:(j',i_{j'})\:(1 \!\leq\! i_j \!\leq\! k_j,\: 
1 \!\leq\! i_{j'} \!\leq\! k_{j'})\) are the row, column index respectively.
Denoting by \(X^{\vee}_{\rm cusp}|_{x^o}\) the variety of the Hessian 
\(\det y \!=\!0\) in the \(x^o\)-section \(X^{\vee}|_{x^o}\) of 
Eq.~(\ref{eq:x^o}), we can define \(X^{\vee}_{\rm cusp}\) as 
\begin{equation}
\label{eq:cusp}
X^{\vee}_{\rm cusp} = X^{\vee}_{\rm cusp}|_{x^o}\cdot G.
\end{equation}
This \(X^{\vee}_{\rm cusp}\) is already closed without taking the closure.

\section{Classification of multipartite entanglement}
\label{sec:4}
\noindent
According to Sec.2 and Sec.3, we illustrate the classification
of multipartite pure entangled states for typical cases.

\subsection{3-qubit (format \(2^3\)) case}
\label{sec:3qubit}
\begin{figure}[t]
\begin{center}
\begin{psfrags}
\psfrag{m}{\small $M$}
\psfrag{v}{\small $X^{\vee}$}
\psfrag{s}{\small $X^{\vee}_{\rm cusp}$}
\psfrag{a}{\color{red}\small $X^{\vee}_{\rm node}(1)$ }
\psfrag{b}{\color{blue}\small $X^{\vee}_{\rm node}(2)$}
\psfrag{c}{\color{dgreen}\small $X^{\vee}_{\rm node}(3)$}
\psfrag{x}{\small $X$}
\psfrag{G}{\rotatebox{60}{GHZ}}
\psfrag{W}{\rotatebox{60}{W}}
\includegraphics[clip]{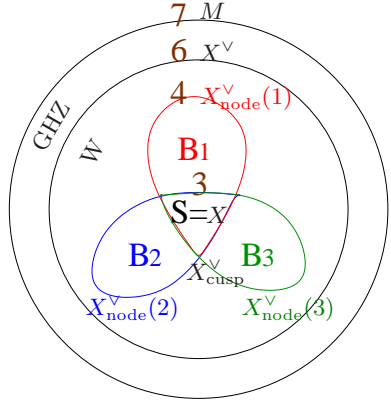}
\caption{
The onion-like classification of SLOCC orbits in the \(3\)-qubit case.
We utilize a duality between the smallest closed subvariety 
\(X\) and the largest closed subvariety \(X^{\vee}\).
The dual variety \(X^{\vee}\) (zero hyperdeterminant) and 
its singularities constitute SLOCC-invariant closed subvarieties 
so that they classify the multipartite entangled states (SLOCC orbits).}
\label{fig:onion_3bit}
\end{psfrags}
\end{center}
\end{figure}

\noindent
The classification of the 3 qubits under SLOCC has been already done in 
\cite{dur+00,note1}.
Surprisingly, Gelfand {\it et al.} considered the same mathematical problem 
by \({\rm Det}A_3\) in Example 4.5 of \cite{gelfand+94}. 
Our idea is inspired by this example.
We complement the Gelfand {\it et al.}'s result, analyzing additionally  
the singularities of \(X^{\vee}\) in details. 
The dimensions, representatives, names, and varieties of the orbits are 
summarized as follows. 
The basis vector \(|i_1\rangle\otimes | i_2 \rangle\otimes |i_3\rangle\) 
is abbreviated to \(|i_1 i_2 i_3 \rangle\).
\medskip 

\noindent
dim 7:  \(|000\rangle + |111\rangle,\)       GHZ  
        \(\in\: M(={\mathbb C}P^7)\!-\!X^{\vee}\). \\
dim 6:  \(|001\rangle + |010\rangle + |100\rangle,\) W  
        \(\in\: X^{\vee}\!-\!X^{\vee}_{\rm sing}
        =X^{\vee}\!-\!X^{\vee}_{\rm cusp}\). \\
dim 4:  \(|001\rangle + |010\rangle,\: |001\rangle + |100\rangle,\: 
        |010\rangle + |100\rangle,\)  
        biseparable \(B_{j}\) \(\in\: X^{\vee}_{\rm node}(j) \!-\! X\) 
        for \(j=1,2,3\). \\
\qquad\quad\,  
        \(X^{\vee}_{\rm node}(j)={\mathbb C}P^1_{j{\rm \mbox{-}th}} \!\times\! 
        {\mathbb C}P^3\) are three closed irreducible 
        components of \(X^{\vee}_{\rm sing}=X^{\vee}_{\rm cusp}\).\\
dim 3:  \(|000\rangle,\) completely separable \(S\) 
        \(\in\: X= \bigcap_{j=1,2,3} X^{\vee}_{\rm node}(j) 
        ={\mathbb C}P^1 \!\times\!{\mathbb C}P^1 \!\times\!{\mathbb C}P^1\). 
\medskip  

\(G\!=\!GL_2 \times GL_2 \times GL_2\) has the onion structure of six orbits 
on \(M\) (see Fig.~\ref{fig:onion_3bit}), by excluding the orbit 
\(\emptyset \;(\!=\!X^{\vee}_{\rm node}(\emptyset))\).
The dual variety \(X^{\vee}\) is given by \({\rm Det}A_3 =0\) 
(cf. Eq.~(\ref{eq:tau})). 
Its dimension is \(7-1=6\).
The outside of \(X^{\vee}\) is generic tripartite entangled class of 
the maximal dimension, whose representative is GHZ.
This suggests that almost any state in the \(3\) qubits can be locally 
transformed into GHZ with a finite probability, and vice versa. 
Next, we can identify \(X^{\vee}_{\rm sing}\) as \(X^{\vee}_{\rm cusp}\), 
which is the union of three closed irreducible subvarieties 
\(X^{\vee}_{\rm node}(j)\) for \(j\!=\!1,2,3\) (cf. \cite{weyman+96}). 
For example, \(X^{\vee}_{\rm node}(1)\) means by definition that, 
in addition to the condition for \(X^{\vee}\) in Sec.2,
there exists some nonzero \(x^{(1)}\) such that \(F(A,x)\!=\!0\)  
for any \(x^{(2)},x^{(3)}\); i.e., a set of linear equations 
\begin{equation}
y_{i_2,i_3}(x^{(1)}) =
\frac{\partial^2}{\partial x^{(2)}_{i_2}\partial x^{(3)}_{i_3}}F(A,x) = 0
\quad \mbox{for} \quad
i_j = 0,1 
\end{equation}
has a nontrivial solution \(x^{(1)}\).
This indicates that not only \(X^{\vee}_{\rm node}(1) \subset 
X^{\vee}_{\rm cusp}\), but the "bipartite" matrix
\begin{equation}
\label{eq:B1}
\left(\begin{array}{cccc}
a_{000} & a_{001} & a_{010} & a_{011} \\
a_{100} & a_{101} & a_{110} & a_{111}
\end{array}\right)
\end{equation}
never has the full rank  (i.e., six \(2 \!\times\! 2\) minors in 
Eq.(\ref{eq:B1}) are zero). We can identify \(X^{\vee}_{\rm node}(1)\) as 
the set \({\mathbb C}P^{1}_{1{\rm st}}\times{\mathbb C}P^{3}\), seen in 
Sec.2, of biseparable states between the 1st party and 
the rest of the parties.
Its dimension is \(1+3=4\).
Likewise, \(X^{\vee}_{\rm node}(j)\) for \(j=2,3\) gives the biseparable
class for the 2nd, 3rd party, respectively.
So, the class of \(X^{\vee}\!-\! X^{\vee}_{\rm sing}\) is found to be 
tripartite entangled states, whose representative is W.
We can intuitively see that, among genuine tripartite entangled states, 
W is rare, compared to GHZ \cite{dur+00}.
Finally, the intersection of \(X^{\vee}_{\rm node}(j)\) is the completely 
separable class \(S\), given by the Segre variety \(X\) of dimension \(3\).
Another intuitive explanation about this procedure is seen in the appendix
of \cite{miyake02}.

Now we clarify the relationship of six classes by {\it noninvertible} local
operations.
Because noninvertible local operations cause the decrease in local ranks 
\cite{note5}, the partially ordered structure of entangled states in the 
\(3\) qubits, included in Fig.~\ref{fig:part_order}, appears.
Two inequivalent tripartite entangled classes, GHZ and W, have the same local 
ranks \((2,2,2)\) for each party so that they are not interconvertible 
by the noninvertible local operations (i.e., general LOCC).
Two classes hold different physical properties \cite{dur+00}; the GHZ 
representative state has the maximal amount of generic tripartite entanglement
measured by the \(3\)-tangle \(\tau \!=\! 4|{\rm Det}A_3| \), while 
the W representative state has the maximal amount of (average) \(2\)-partite 
entanglement distributed over \(3\) parties (also \cite{koashi+00}). 
Under LOCC, a state in these two classes can be transformed into any state 
in one of the three biseparable classes \(B_j \;(j=1,2,3)\), where
the \(j\)-th local rank is \(1\) and the others are \(2\). 
Three classes \(B_j\) never convert into each other.
Likewise, a state in \(B_j\) can be locally transformed into any state 
in the completely separable class \(S\) of local ranks \((1,1,1)\).

This is how the onion-like classification of SLOCC orbits reveals that 
multipartite entangled classes constitute the partially ordered structure.
It indicates significant differences from the totally ordered one in the 
bipartite case.
(i) In the \(3\)-qubit case, all SLOCC invariants we need to classify is 
the hyperdeterminant \({\rm Det}A_3\) in addition to local ranks.
(ii) Although noninvertible local operations generally mean
the transformation into the further inside of the onion structure, 
an outer class can not necessarily be transformed into the {\it neighboring} 
inner class. A good example is given by GHZ and W, as we have just seen.

%
\subsection{Format \({\rm 3\times 2\times 2} \) case}
\label{sec:3x2x2}
\noindent
Before proceeding the \(n\!\geq\! 4\)-qubit case, we drop in the format 
\(3 \!\times\! 2 \!\times\! 2\), which would give an insight into 
the structure of multipartite entangled states when each party has 
a system consisted of more than two levels. 
This case is interesting since on the one hand (contrary to the \(3\)-qubit 
case), it is typical that GHZ and W are included in \(X^{\vee}_{\rm sing}\); 
on the other hand (similarly to the bipartite or \(3\)-qubit cases),
SLOCC has still finite classes so that it becomes another good test for the 
equivalence to the SLOCC classification.
Besides, it is a boundary format so that several subvarieties can be 
explicitly calculated, and enables us to analyze entanglement in the 
qubit-system using an auxiliary level, like ion traps.  
\medskip

\noindent
dim 11: \(|000\rangle + |101\rangle + |110\rangle + |211\rangle\) 
        \(\in\: M(={\mathbb C}P^{11})\!-\!X^{\vee}\). \\
dim 10: \(|000\rangle + |101\rangle + |211\rangle\)
        \(\in\: X^{\vee}\!-\!X^{\vee}_{\rm sing}
        =X^{\vee}\!-\!X^{\vee}_{\rm node}(1)\).\\  
dim 9:  \(|000\rangle + |111\rangle\), GHZ 
        \(\in\: X^{\vee}_{\rm sing}(\!=\!X^{\vee}_{\rm node}(1))
                \!-\!X^{\vee}_{\rm cusp}\). \\
dim 8:  \(|001\rangle + |010\rangle + |100\rangle\), W   
        \(\in\: X^{\vee}_{\rm cusp} \!-\! 
          \bigcup_{j=\emptyset,2,3}X^{\vee}_{\rm node}(j) \).\\
dim 6:  \(|001\rangle + |100\rangle,\: |010\rangle + |100\rangle,\) 
        biseparable \(B_2,\: B_3\)
        \(\in\: X^{\vee}_{\rm node}(2) \!-\! X,\:
                X^{\vee}_{\rm node}(3) \!-\! X\). \\
dim 5:  \(|001\rangle + |010\rangle\), biseparable \(B_1\) 
        \(\in\: X^{\vee}_{\rm node}(\emptyset) \!-\! X\). \\
dim 4:  \(|000\rangle\), completely separable \(S\) 
        \(\in X = {\mathbb C}P^{2}\!\times{\mathbb C}P^{1}
        \!\times{\mathbb C}P^{1}\). 
\medskip

\begin{figure}[tb]
\begin{center}
\begin{psfrags}
\psfrag{m}{\small $M$}
\psfrag{v}{\small $X^{\vee}$}
\psfrag{g}{\small $X^{\vee}_{\rm node}(1)$}
\psfrag{w}{\small $X^{\vee}_{\rm cusp}$}
\psfrag{a}{{\color{red}\small $X^{\vee}_{\rm node}(\emptyset)$}}
\psfrag{b}{{\color{blue}\small $X^{\vee}_{\rm node}(2)$}}
\psfrag{c}{{\color{dgreen}\small $X^{\vee}_{\rm node}(3)$}}
\psfrag{x}{\small $X$}
\psfrag{F}{\rotatebox{60}
{$|000\rangle\!+\!|101\rangle\!+\!|110\rangle\!+\!|211\rangle$}}
\psfrag{T}{\rotatebox{60}{$|000\rangle\!+\!|101\rangle\!+\!|211\rangle$}}
\psfrag{G}{\rotatebox{60}{GHZ}}
\psfrag{W}{\rotatebox{60}{W}}
\includegraphics[clip]{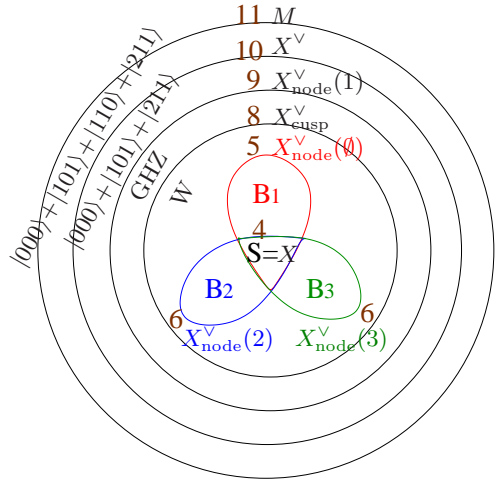}
\caption{ 
The onion-like classification of SLOCC orbits in the 
\(3 \!\times\! 2 \!\times\! 2\) format.
Although this resembles Fig.~\ref{fig:onion_3bit} in the order of SLOCC 
orbits (two orbits are added outside), it is worth while to note that 
singularities of \(X^{\vee}\), which classify the SLOCC orbits, have 
a different order. } 
\label{fig:onion_3x2x2}
\end{psfrags}
\end{center}
\end{figure}
\begin{figure}[tb]
\begin{center}
\includegraphics[clip]{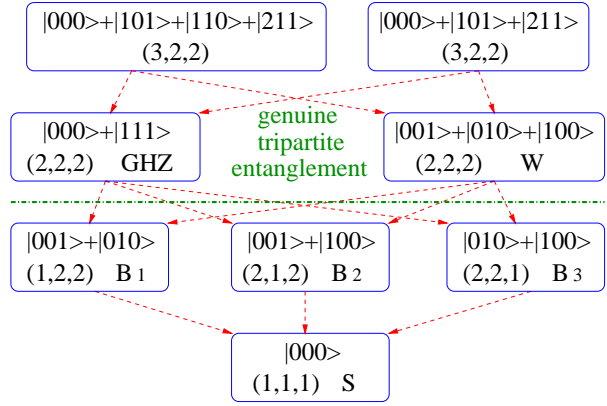}
\caption{
The partially ordered structure of multipartite pure entangled states
in the \(3\times 2\times 2\) format, including the \(3\)-qubit case.
Each class, corresponding to the SLOCC orbit, is labeled by 
the representative, local ranks, and the name.
Noninvertible local operations, indicated by dashed arrows, 
degrade "higher" entangled classes into "lower" entangled ones.}  
\label{fig:part_order}
\end{center}
\end{figure}

The onion structure consists of eight orbits on \(M\) under SLOCC 
(see Fig.~\ref{fig:onion_3x2x2}).
Generic entangled states of the outermost class is given by nonzero 
\({\rm Det}A\), which can be calculated in the {\it boundary} format as the 
determinant associated with the Cayley-Koszul complex. 
Although this is one of the Gelfand {\it et al.}'s recent successes for 
generalized discriminants, we avoid its detailed explanation here.
According to Theorem 3.3 of \cite{gelfand+94}, we have 
\begin{equation}
\label{eq:DetA_3x2x2}
{\rm Det}A = m_1 m_4 -m_2 m_3
\end{equation}
of degree \(6\), where \(m_j \;(j=1,2,3,4)\) is the \(3 \times 3\) minor of 
\begin{equation}
\label{eq:A_3x2x2}
\left( \begin{array}{cccc}
a_{000} & a_{001} & a_{010} & a_{011} \\
a_{100} & a_{101} & a_{110} & a_{111} \\
a_{200} & a_{201} & a_{210} & a_{211}
       \end{array} \right) 
\end{equation}
without the \(j\)-th column, respectively.
Next, it is characteristic that \(X^{\vee}_{\rm sing}\) is 
\(X^{\vee}_{\rm node}(1)\) \cite{weyman+96}. 
Similarly to the \(3\)-qubit case in Sec.~4.1, 
\(X^{\vee}_{\rm node}(1)\) means that the "bipartite" matrix in 
Eq.~(\ref{eq:A_3x2x2}) does not have the full rank, i.e., all four 
\(3\times 3\) minors \(m_j\) in Eq.~(\ref{eq:A_3x2x2}) are zero.
The SLOCC orbits which appear inside \(X^{\vee}_{\rm sing}\) are essentially 
the same as the 3-qubit case.

Accordingly, we obtain the partially ordered structure of multipartite 
entangled states as Fig.~\ref{fig:part_order}.
The tripartite entanglement consists of four classes.
Because the class of \(M\!-\! X^{\vee}\), whose representative is 
\(|000\rangle \!+\!|101\rangle \!+\!|110\rangle \!+\!|211\rangle\), 
and that of \(X^{\vee}\!-\! X^{\vee}_{\rm sing}\), whose representative is 
\(|000\rangle \!+\!|101\rangle \!+\!|211\rangle\), have the same local ranks 
\((3,2,2)\), they do not convert each other in the same reason as GHZ and W 
do not.  
However, the former two classes of the local ranks \((3,2,2)\) can convert to 
the latter two classes of \((2,2,2)\) by noninvertible local operations 
(i.e., LOCC). And we can "degrade" these tripartite entangled classes into 
the biseparable or completely separable classes by LOCC in a similar fashion 
to the \(3\) qubits.

We notice that 3 grades in the \(3\)-qubit case changed to 4 grades in 
the \(3 \!\times\! 2 \!\times\! 2\) (\(1\)-qutrit and \(2\)-qubit) case.
In general, the partially ordered structure becomes "higher",
as the system of each party becomes the higher dimensional one. 
We also see how the tensor rank \cite{note3} is inadequate for the 
onion-like classification of SLOCC orbits.

%
\subsection{\(n\!\geq\!4\)-qubit (format \(2^n\)) case}
\label{sec:nqubit}
\noindent
Further in the \(n\!\geq\!4\)-qubit case, our classification works. 
The outermost class \(M(\!=\!{\mathbb C}P^{2^{n}\!-\!1})\!-\!X^{\vee}\) of 
generic \(n\)-partite entangled states is given by \({\rm Det}A_n \!\ne\! 0\). 
In \(n \!=\! 4\), \({\rm Det}A_4\) of degree 24 is explicitly calculated 
by the Schl\"afli's construction in Sec.3.2.
It would be suggestive to transform any generic \(4\)-partite state 
(\({\rm Det}A_4 \!\ne\! 0\)) to the "representative" of the outermost class 
by invertible local operations,
\begin{equation}
\label{eq:generic_4bit}
\alpha(|0000\rangle +|1111\rangle)
+\beta(|0011\rangle+|1100\rangle) 
+\gamma(|0101\rangle +|1010\rangle)
+\delta(|0110\rangle+|1001\rangle),
\end{equation}
where the continuous complex coefficients \(\alpha,\beta,\gamma,\delta\) 
should satisfy
\begin{multline}
\label{eq:det4}
{\rm Det}A_4 = \alpha^2 \beta^2 \gamma^2 \delta^2 
(\alpha+\beta+\gamma+\delta)^2 (\alpha+\beta+\gamma-\delta)^2  
(\alpha+\beta-\gamma+\delta)^2 (\alpha-\beta+\gamma+\delta)^2 \\
(-\alpha+\beta+\gamma+\delta)^2 
(\alpha+\beta-\gamma-\delta)^2 (\alpha-\beta+\gamma-\delta)^2 
(\alpha-\beta-\gamma+\delta)^2 \ne 0.  
\end{multline}
Thus three complex parameters remain in the outermost class (since we 
consider rays rather than normalized state vectors).
This means that there are infinitely many same dimensional SLOCC orbits 
in the \(4\) qubits, and the SLOCC orbits never locally convert to each other
when their sets of the parameters are distinct.
It is also the case for the \(n \!>\! 4\) qubits.
Note that, in \(n\!=\!4\), this outermost class \(M\!-\!X^{\vee}\) corresponds
to the family of generic states in Verstraete {\it et al.}'s classification 
of the \(4\) qubits by a different approach (generalizing
the singular value decomposition in matrix analysis to complex 
orthogonal equivalence classes), and \(X^{\vee}\) contains their 
other special families \cite{verstraete+02}. 

The next outermost class is \(X^{\vee}\!-\!X^{\vee}_{\rm sing}\). 
In the \(4\) qubits, \(X^{\vee}_{\rm sing}\) is shown to consist of eight 
closed irreducible components of codimension 1 in \(X^{\vee}\); 
\(X^{\vee}_{\rm cusp},\: X^{\vee}_{\rm node}(\emptyset)\), and six 
\(X^{\vee}_{\rm node}(j_1,j_2)\) for \(1 \!\leq\! j_1 \!<\! j_2 \!\leq\! 4\) 
\cite{weyman+96}.
They neither contain nor are contained by each other.
Their intersections also give (finitely) many lower dimensional genuine
\(4\)-partite entangled classes. 
Since the \(4\)-partite entangled classes necessarily have the same local 
ranks \((2,2,2,2)\), these classes are not interconvertible by noninvertible 
local operations (i.e., any LOCC).
As typical examples, GHZ, the maximally entangled state in Bell's 
inequalities \cite{gisin+98}, 
\begin{equation}
|{\rm GHZ}\rangle = |0000\rangle + |1111\rangle,
\end{equation}
(i.e., \(a_{0000}=a_{1111} \ne 0\) and the others are \(0\)) is included in 
the intersection of \(X^{\vee}_{\rm node}(\emptyset)\) and six 
\(X^{\vee}_{\rm node}(j_1,j_2)\), but is excluded from \(X^{\vee}_{\rm cusp}\).
In contrast, W, 
\begin{equation}
|{\rm W}\rangle = |0001\rangle + |0010\rangle + |0100\rangle + |1000\rangle,
\end{equation}
(i.e., \(a_{0001}=a_{0010}=a_{0100}=a_{1000} \ne 0\) and the others are \(0\)) 
is included in the intersection of \(X^{\vee}_{\rm cusp}\) and  
six \(X^{\vee}_{\rm node}(j_1,j_2)\) but is excluded from 
\(X^{\vee}_{\rm node}(\emptyset)\).  

In the \(n\!>\!4\) qubits, \(X^{\vee}_{\rm sing}\) is shown to consist of 
just two closed irreducible components \(X^{\vee}_{\rm cusp}\) and 
\(X^{\vee}_{\rm node}(\emptyset)\) \cite{weyman+96}.
We find that GHZ and W are contained not only in 
\(X^{\vee}\:({\rm Det}A_n\!=\!0)\) but in \(X^{\vee}_{\rm sing}\); i.e.,
they have nontrivial solutions in Eqs.(\ref{eq:critical}), satisfying the
singular conditions.
They correspond to different intersections of further singularities similarly 
to the \(4\) qubits. In other words, they are peculiar, living in the border 
dimensions between entangled states and separable ones, 

In brief, the dual variety \(X^{\vee}\) and its singularities lead to 
the {\it coarse} onion-like classification of SLOCC orbits, when SLOCC gives
infinitely many orbits.
The partially ordered structure of multipartite pure entangled states
becomes "wider", as the number \(n\) of parties increases.
Although many inequivalent \(n\)-partite entangled classes appear in the 
\(n\) qubits, they never locally convert to each other, as observed in 
\cite{dur+00}.
In particular, the majority of the \(n\)-partite entangled states never 
convert to GHZ (or W) by LOCC, and the opposite conversion is also
not possible. 
This is a significant difference from the bipartite or \(3\)-qubit case,
where almost any entangled states and the maximally entangled states 
(GHZ) can convert to each other by LOCC with nonvanishing probabilities.

%
\section{Conclusion}
\label{sec:conclude}
\noindent
We have classified multipartite entanglement (SLOCC orbits) in a unified 
manner based on hyperdeterminants \({\rm Det}A\).
The underlying idea is the duality between the set of completely separable 
states (the Segre variety \(X\)) and that of degenerate entangled states
(its dual variety \(X^{\vee}\) of \({\rm Det}A \!=\! 0\)).
The generic entangled class of the maximal dimension is given by the outside
of \(X^{\vee}\), and other multipartite entangled classes appear in
\(X^{\vee}\) or its different singularities, seen in the onion picture
like Fig.~\ref{fig:onion_3bit} or Fig.~\ref{fig:onion_3x2x2}.
Since the onion-like classification of SLOCC orbits is given by every closed 
subset, not only it is useful to see intuitively why, say in 
the \(3\) qubits, the W class is rare compared to the GHZ class, but 
it can be also extended to the classification of multipartite mixed states
(cf. the appendix of \cite{miyake02}).

In virtue of this onion-like classification, we clarify the partially ordered 
structure, such as Fig.~\ref{fig:part_order}, of inequivalent multipartite 
entangled classes of pure states, which is significantly different from the 
totally ordered one in the bipartite case.
Local ranks are not enough to distinguish these classes any more, and 
we need to calculate SLOCC invariants associated with \({\rm Det}A\).
This partially ordered structure becomes "higher" as the dimensions of 
subsystems enlarge, and it becomes "wider" as the number of the parties 
increases.

This work reveals that the situation of the widely known bipartite or 
\(3\)-qubit cases, where the maximally entangled states in Bell's 
inequalities belong to the generic class, is exceptional.
Lying far inside the onion structure, the maximally entangled states (GHZ) 
are included in the lower dimensional peculiar class in general, e.g., for 
the \(n \geq 4\) qubits.
It suggests two points.
The majority of multipartite entangled states can not convert to GHZ by LOCC, 
and vice versa. So, we have given an alternative explanation to this 
observation, first made in \cite{dur+00} by comparing the number of local 
parameters accessible in SLOCC with the dimension of the whole Hilbert space.
Moreover, there seems no a priori reason why we choose GHZ states as
the {\it canonical} \(n\)-partite entangled states, which, for example,
constitute a minimal reversible entanglement generating set (MREGS) in
asymptotically reversible LOCC \cite{bennett+00,wu+00}.

The onion-like classification seems to be reasonable in the sense that
it coincides with the SLOCC classification when SLOCC gives finitely many 
orbits, such as the bipartite or \(3\)-qubit cases.
So two states belonging to the same class can convert each other by 
invertible local operations with nonzero probabilities.
On the other hand, when SLOCC gives infinitely many orbits, 
this classification is still SLOCC-invariant, but may contain in one class 
infinitely many {\it same} dimensional SLOCC orbits which can not locally 
convert to each other even probabilistically.
For example, in the \(4\)-qubit case, the generic entangled class in 
Eq.(\ref{eq:generic_4bit}) has three nonlocal continuous parameters.
Note that it can be possible to make the onion-like classification finer, by 
characterizing the nonlocal continuous parameters in each class.

Then, we may ask, what is the physical interpretation of the onion-like 
classification in the case of infinitely many SLOCC orbits? 
Although a simple answer has yet to be found, we discuss two points.
(i) Let us consider {\it global} unitary operations which create 
the multipartite entanglement.
On the one hand, states in distinct classes would have the different 
complexity of the global operations, since they have the distinct number 
and pattern of nonlocal parameters.
On the other hand, states in one class are supposed to have the equivalent
complexity, since they just correspond to different "angles" of the global 
unitary operations.
(ii) We can consider the case where {\it more than one} states are shared,
including the asymptotic case.
Even in two shared states, there can exist a local conversion which is 
impossible if they are operated separately, such as the catalysis 
effect \cite{jonathan+99}. 
So we can expect that we do more locally in this situation and the coarse
classification may have some physical significance.
This problem remains unsettled even in the bipartite case. 

Finally, three related topics are discussed. 
(i) The absolute value \(|{\rm Det}A_n|\) of the hyperdeterminant, 
representing the amount of generic entanglement, is an entanglement monotone 
by Vidal \cite{vidal00}.
This never conflicts with the property that the maximally entangled states
in Bell's inequalities (GHZ) generally has a zero \({\rm Det}A_n\). 
A single entanglement monotone is insufficient to judge the LOCC 
convertibility, and generic entangled states of the nonzero \({\rm Det}A_n\) 
can not convert to GHZ in spite of decreasing \(|{\rm Det}A_n|\).
(ii) The \(3\)-tangle \(\tau = 4|{\rm Det}A_3|\) first appeared in the 
context of so-called entanglement sharing \cite{coffman+00}; i.e., 
in the \(3\) qubits, there is a constraint (trade-off) between the amount of 
\(2\)-partite entanglement and that of \(3\)-partite entanglement.
By using the entanglement measure (concurrence \(C\)) for the \(2\)-qubit 
{\it mixed} entangled states, this is written as 
\(C^2_{1(23)} \geq C^2_{12} + C^2_{13}\), and \(\tau\) is defined by 
\(\tau = C^2_{1(23)}-C^2_{12}-C^2_{13}\) for the \(3\)-qubit {\it pure} 
entangled states.
We expect that, in turn, the hyperdeterminant \({\rm Det}A_n\) gives a clue 
to find the entanglement measure of more than \(2\)-qubit {\it mixed} states.
(iii) In the classification of mixed states, we can construct the so-called
witness operator \({\mathcal W}\) in order to detect the entanglement of 
a given mixed state \(\rho\) \cite{lewenstein+00}.
It would be interesting to observe that since the optimal one 
\({\mathcal W}_{o}\) forms the tangent hyperplane 
\({\rm tr}(\rho {\mathcal W}_{o})=0\) to the set of separable mixed states, 
it shares the same ideas as our dual variety. 

We hope that many intrinsic features of multipartite entanglement 
will be elucidated from hyperdeterminants.

{\it Note added.} 
For the 4-qubit case, a complete, generating set for polynomial invariants 
under SLOCC is calculated recently in \cite{luque+02}, which enables
\({\rm Det}A_4\) expressed by the lower degree invariants.

\section*{Acknowledgments}
\noindent
One of the authors (A.M.) would like to thank the participants of the 
ERATO workshop on Quantum Information Science (September 5-8, 2002, Tokyo, 
Japan) for the most helpful and enjoyable discussions.
The support by the ERATO Quantum Computation and Information Project
is also acknowledged.



\begin{thebibliography}{000}
\bibitem{note1}
SLOCC was tied to \(SL \times SL\) in \cite{dur+00}.
Since we treat \(|\Psi\rangle\) as a ray, i.e., an unnormalized vector, we 
rather relate SLOCC to \(GL \times GL\).
\bibitem{note2}
The local rank can be defined as the rank of the reduced density matrix 
traced out for all except one party.
Note that this definition is applicable to the multipartite case.
\bibitem{note4}
For example, let us consider two Einstein-Podolsky-Rosen (EPR) pairs
\(\frac{1}{\sqrt{2}}(|00\rangle +|11\rangle)_{12}\otimes 
\frac{1}{\sqrt{2}}(|00\rangle + |11\rangle)_{34}\) in the \(4\) qubits.
Since their local ranks are \((2,2,2,2)\), we can not distinguish
this state from genuine \(4\)-partite entangled states 
(cf. Sec.4.3). In contrast, we readily find that the state is 
included in \({\mathbb C}P^{3} \!\times\! {\mathbb C}P^{3}\) so that it is 
separable (not genuine \(4\)-partite entangled).
\bibitem{note5}
If there exists a noninvertible local operation between SLOCC orbits, 
some of local ranks decrease. 
Note that the converse is generally not the case.
However in our examples in Sec.4, we do find some noninvertible
local operations for the decrease of local ranks.
\bibitem{note3}
The tensor rank means the number of terms in the minimal decomposition of 
\(|\Psi\rangle\) in Eq.(\ref{eq:multi}) by separable states.
It coincides with the local rank in the bipartite case.
\end{thebibliography}
\end{document}